\documentclass[a4paper,reqno,pre,superscriptaddress,twocolumn]{revtex4}

\usepackage{graphicx,color}
\usepackage{dcolumn}
\usepackage{epsfig}

\usepackage[centertags]{amsmath}
\usepackage{amsfonts,amsmath}
\usepackage{euscript}
\usepackage{amssymb}
\usepackage{amsthm}
\usepackage{newlfont}
\usepackage{mathrsfs}
\usepackage{subfigure}

\newcommand{\opunit}{\text{1}\kern-0.22em\text{l}}

\newcommand{\id}{\textrm{d}}

\def\bea{\begin{eqnarray}}
\def\eea{\end{eqnarray}}
\def\ba{\begin{array}}
\def\ea{\end{array}}
\def\n{\nonumber}

\def\la{\langle}
\def\ra{\rangle}

\begin{document}
 
\title{Active Brownian Motion in Two Dimensions}
\author{Urna Basu}
\affiliation{LPTMS, CNRS, Univ. Paris-Sud, Universit\'{e} Paris-Saclay, 91405 Orsay, France}
\affiliation{Raman Research Institute, Bangalore 560080, India}
\author{Satya N. Majumdar}
\affiliation{LPTMS, CNRS, Univ. Paris-Sud, Universit\'{e} Paris-Saclay, 91405 Orsay, France}
\author{Alberto Rosso}
\affiliation{LPTMS, CNRS, Univ. Paris-Sud, Universit\'{e} Paris-Saclay, 91405 Orsay, France}
\author{Gr\'egory Schehr}
\affiliation{LPTMS, CNRS, Univ. Paris-Sud, Universit\'{e} Paris-Saclay, 91405 Orsay, France}
\pacs{05.70.Ln %Nonequilibrium and irreversible thermodynamics
05.40.-a %Fluctuation phenomena, random processes, noise, and Brownian motion
83.10.Pp %particle dynamics
}

\begin{abstract}
%Active Brownian self-propell themselves along a direction which itself diffuses slowly. 

We study the dynamics of a single active Brownian particle (ABP) in two spatial dimensions. The ABP has an intrinsic time scale $D_R^{-1}$ set by the rotational diffusion constant $D_R$. We show that, at short-times $t \ll D_R^{-1}$, the presence of `activness' results in a strongly anisotropic and non-diffusive dynamics in the $(xy)$ plane. We compute exactly the marginal distributions of the $x$ and $y$ position coordinates along with the radial distribution, which are all shown to be non-Brownian. 
In addition, we show that, at early times, the ABP has anomalous first-passage properties, characterized by non-Brownian exponents. 
\end{abstract}

\maketitle

\section{Introduction}

Active particles form a class of nonequilibrium systems which are able to generate dissipative directed motion through self-propulsion and consuming energy from their environment \cite{Romanczuk,soft, BechingerRev,Ramaswamy2017,Marchetti2017,Schweitzer}. Study of active particles is relevant in a wide variety of biological and soft matter systems ranging from  bacterial motion \cite{Berg2004, Cates2012}, cellular tissue behavior \cite{tissue}, formation of fish schools \cite{Vicsek, fish} as well as granular matter \cite{gran1,gran2} and colloidal surfers \cite{cluster2}. Recent years have seen a tremendous surge of research, both theoretical and experimental, on active matter e.g., the collective behavior of active particles which include flocking \cite{flocking1, flocking2}, clustering \cite{cluster1,cluster2,evans}, phase separation \cite{separation1, separation2, separation3} and the absence of a well defined pressure \cite{Tailleur2015}.

Remarkably, even at a single particle level, active particles show many interesting features like anomalous dynamical behavior \cite{Dhar2017, Erdman1, Erdman2}, non-Boltzmann stationary distribution \cite{RTP, RTP2,Erdman1,Erdman2, Das, Maggi2014,Maggi2015} and accumulation near confining boundaries \cite{boundary, boundary1, boundary2}. 
%Various models of active particles have been studied in literature, which share the common feature ...
One of the simplest and most extensively studied models of active particles is the so called active Brownian particle (ABP) \cite{ABP,BechingerRev,Marchetti2017,Potosky2012,Solon2015} which describes directed spatial motion of overdamped particles at a fixed speed with the direction performing a rotational diffusion, with diffusion constant $D_R$. Interestingly, the same model was also studied as a toy model of computer vision~\cite{Mumford}, as well as in reaction-diffusion systems~\cite{Gredat}. Despite the apparent simplicity of the model, an exact analytical description of the dynamics, beyond the mean-squared radial displacement~\cite{BechingerRev,Sevilla,Howse}, is unfortunately still lacking.

% 
% In a recent experiment, Janus swimmers were confined in a two-dimensional harmonic-like trap with the use of an acoustic tweezer and the stationary density was measured by varying the trap strength~\cite{Takatori}. Strong signatures of activity were observed
% even in the dilute limit, with a crossover from a Gaussian-like stationary state, to a strongly active stationary state, where the particles cluster at the outskirts of the trap. The dilute limit corresponds to a collection of non-interacting Active Brownian Particles (ABP), each one  performing an overdamped directed spatial motion at a fixed speed but with the direction undergoing a rotational diffusion \cite{ABP}. Numerical studies of a single ABP, in presence of confining potentials, have also observed a similar crossover in the stationary state \cite{Solon2015,Potosky2012}. 

The diffusion of the rotational degree of freedom sets a time scale $D_R^{-1}$ which characterizes the persistence of the direction for an ABP. At very late times, one indeed recovers Brownian diffusion with an effective diffusion constant~\cite{BechingerRev,Marchetti2017}. However, this effective Brownian picture does not hold at short times where memory effects are important. In this Letter, we show that at {\it short times} the ABP exhibits a strikingly different behavior compared to the ordinary Brownian motion, with clear fingerprints of `activeness' of the motion. For a given initial orientation of the velocity, we show that at short-times the dynamics is highly anisotropic, with the typical displacements along and transverse to the initial orientation scaling very differently with time. The corresponding distributions turn out to be strongly non-Brownian in nature. We compute exactly these marginal position distributions at short times using path-integral techniques for Brownian functionals. We show that, at short times, the dynamics transverse to the initial orientation can be mapped to the ``Random Acceleration Process'' (RAP)~\cite{Burkhardt2007,satya_review,Bray,Burkhardtbook,Masoliver,Burkhardt,RAP}, a well studied non-Markovian process. Consequently, we show that the ABP exhibits anomalous first-passage properties at short times, with an associated nontrivial exponent in the transverse~direction.

\section {Model} 

We consider an active overdamped particle in the $2d$-plane moving with a constant speed $v_0.$
In addition to its position coordinates $(x,y)$ the particle has an `active' internal degree of freedom, given by the orientational angle $\phi(t)$ of its velocity, which undergoes rotational diffusion. The time evolution is encoded in the Langevin equation~\cite{BechingerRev,Ramaswamy2017,Marchetti2017} 
\begin{subequations}
\bea
\dot x &=&  v_0 \cos \phi(t) \label{eq:modelx}\\
\dot y &=&  v_0 \sin \phi(t)  \label{eq:modely} \\
\dot \phi &=& \sqrt{2D_R}~\eta_\phi(t). \label{eq:phi}
\eea \label{eq:model}
\end{subequations}
Here $\eta_{\phi}(t)$ is a Gaussian white noise with zero mean and correlator $\la \eta_\phi(t) \eta_\phi(t')\ra = \delta(t-t')$ and  
$D_R$ is the associated rotational diffusion constant. The `activeness' in this model stems from the velocity in the $x$ and $y$ directions that are coupled to the orientation $\phi(t)$. This is in contrast to the standard Brownian motion (SBM) where the $x$ and $y$ coordinates undergo uncorrelated translational diffusion with some diffusion constant $D$: $\dot x = \sqrt{2D}~ \eta_x(t),\; \dot y = \sqrt{2D}~ \eta_y(t)$ where $\eta_{x,y}(t)$ are independent delta-correlated white noises. 

The Langevin equations \eqref{eq:model} can be cast in a form similar to that of ordinary Brownian motion where $\xi_x(t) = v_0 \cos \phi(t)$ and $\xi_y(t) = v_0 \sin \phi(t)$ are the effective noises acting on the ABP. However, these `active' effective noises differ from the usual white-noise on two crucial aspects. First, their magnitude is bounded with $\xi_x(t) \le v_0$ (and similarly for $\xi_y(t)$) at all times $t$.  Secondly, both $\xi_x(t)$ and $\xi_y(t)$ have an auto-correlation function  which decays exponentially for large $t_1, t_2,$ 
\bea\label{eq:correl_noise}
\la \xi_x(t_1) \xi_x(t_2) \ra \simeq \frac{v_0^2}{2} \exp [-D_R |t_1 - t_2|]
\eea
and similarly for $\xi_y(t)$ [see Appendix~\ref{sec:noise} for details]. Moreover, $\xi_x$ and $\xi_y$ are also mutually correlated which makes $x$ and $y$-coordinates correlated in ABP, in contrast to the SBM. Clearly, for times $t \gg D_R^{-1}$, the noise correlation in (\ref{eq:correl_noise}) converges to $\la \xi_x(t_1) \xi_x(t_2) \ra \to 2 D_\text{eff}\, \delta(t_1-t_2)$ with an effective diffusion constant $D_\text{eff} = v_0^2/(2 D_R)$. Hence, for $t \gg D_R^{-1}$, ABP effectively reduces to the SBM~\cite{BechingerRev, Marchetti2017}. However, at short-times, the effective noise $\xi_x(t)$ (and $\xi_y(t)$) is correlated in time [see Eq. (\ref{eq:correl_noise})] and we expect memory effects to be important, giving rise to a strong signature of `activeness'.

\begin{figure}[t]
 \centering
 \includegraphics[width=\linewidth]{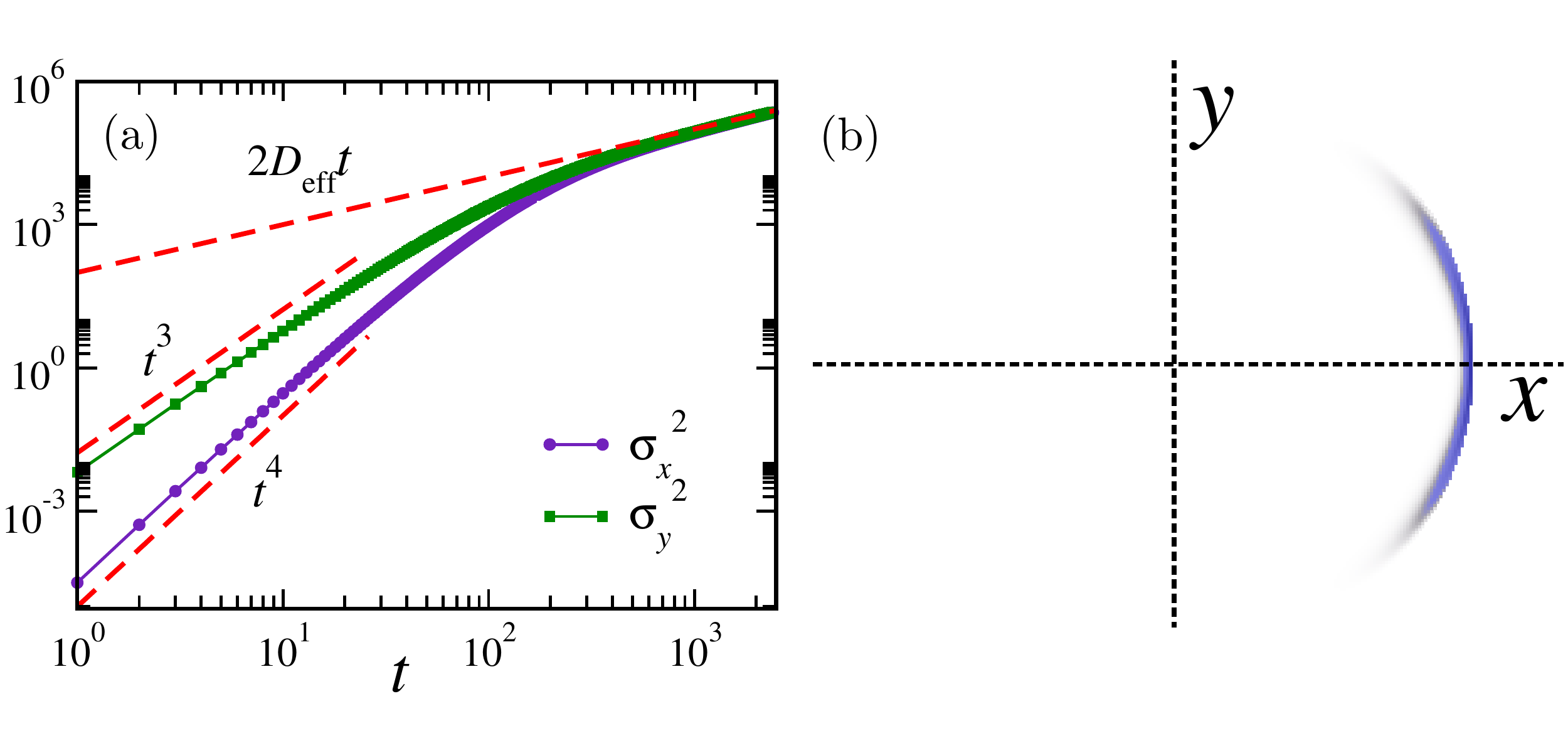} \hspace*{0.05cm} %\includegraphics[width=3.7 cm]{./a.pdf} 
 % AB_cumul.pdf: 638x487 pixel, 72dpi, 22.51x17.18 cm, bb=0 0 638 487
 \caption{(a) Mean-squared displacement $\sigma^2_{x,y}$ of $x$ and  $y$ components of  the position of an ABP as a function of $t$ for $v_0=1$ and $D_R=0.01$. The symbols correspond to simulations and the solid lines are the predictions from the exact calculations. (b) Position probability distribution $P(x,y,t)$ in the $(xy)$ plane for $D_R=0.1$ and $v_0=1$  at an early time $t=1.$  }
 \label{fig:moments}
\end{figure}

% 
% It is convenient to consider the complex coordinate $z(t)=x(t)+i y(t)$ which evolves as
% \bea
% \dot z =  \xi(t) \;\; {\textrm{where}} \;\; \xi(t) = v_0 e^{i \phi(t)} \;. \label{eq:zdot}
% \eea
% The `active' effective noise $\xi(t)$ is very different from the usual white-noise acting on a passive Brownian particle. First, its magnitude is clearly bounded, $|\xi(t)| \le v_0$ at all times $t$. Secondly, $\xi(t)$ has an auto-correlation function [see the Supplemental Material (SM) \cite{SM} for details],
% \bea
% \la \xi(t_1) \bar \xi(t_2) \ra = v_0^2 \exp [-D_R |t_1 - t_2|], \label{eq:expphi}
% \eea
% where $\bar \xi(t) = v_0 e^{-i \phi(t)}$ is the complex conjugate of $\xi(t)$. 

% Clearly, for times longer than the persistence time $\tau_R = D_R^{-1}$, the correlation decays fast, and an effective diffusive behavior can be expected for the ABP. Consequently, the mean-square displacement grows linearly with time, albeit with a renormalized diffusion constant $D_\text{eff} = v_0^2 \tau_R /2$ \cite{BechingerRev, Marchetti2017}. At short-times, however, the noise $\xi(t)$ is strongly correlated and signatures of activity are expected to be visible.\\

\section{Short-time regime}
%\noindent{\it Short-time regime:} 

Consider an ABP starting initially at the origin $x=y=0$ with a given orientation, which we choose, without any loss of generality,  to be along the $x$-axis, so that $\phi(0)=0$. The initial value of $\phi(0)$ selects a specific direction ($x$ here), thereby breaking the symmetry between the dynamics of the $x$ and the $y$ coordinates in Eqs. (\ref{eq:modelx}) and (\ref{eq:modely}). The simplest observable that shows this anisotropy is the mean-squared displacement (MSD) for the $x$ and the $y$ coordinates separately. In fact, the radial MSD in this model $\la r^2\ra = \la x^2 \ra +\la y^2\ra$  was computed long time back~\cite{BechingerRev,Sevilla,Howse}. However, to investigate the anisotropy, we need   
to compute the $x$-MSD $\sigma^2_{x} = \la x^2\ra - \la x\ra^2$ and $y$-MSD $\sigma^2_{y}=\la y^2\ra - \la y\ra^2$ separately. Indeed, in Eq.~\eqref{xy_var_expl} of Appendix \ref{sec:MSD} we show that $\sigma_x^2$ and $\sigma_y^2$ can be computed exactly at all times $t$. From this exact result, the small $t \ll D_R^{-1}$ behavior can be read off, 
\bea
\sigma_x^2 &\approx&  \frac 13 v_0^2 D_R^2 t^4  - \frac 7{15} v_0^2 D_R^3 t^5 + \cdots \\
\sigma_y^2 &\approx&  \frac 23 v_0^2 D_R t^3  - \frac 5{6} v_0^2 D_R^2 t^4 + \cdots \;, \label{eq:xyvar}
\eea
manifesting clearly the anisotropy, as well as the non-diffusive behavior at short times. Evidently, for small $t$, the MSD in the $x$-direction $\sigma_x^2 \sim t^4$ is much smaller than that in the $y$-direction $\sigma_y^2 \sim t^3$.
This is in clear contrast to a SBM where the mean-squared displacements $\sim t$ at all times, and in both $x$ and $y$ directions. Figure \ref{fig:moments} compares the exact result of Eq. \eqref{xy_var_expl} of Appendix \ref{sec:MSD} (solid lines) with the $\sigma^2_{x,y}$ obtained from simulations (symbols). As discussed before, at late times $t \gg D_R^{-1}$, both $\sigma_{x,y}^2 \approx 2 D\,t$ behave diffusively with $D_{\textrm{eff}}=v_0^2/2D_R$ [see Eq. \eqref{xy_var_expl} of Appendix \ref{sec:MSD}]. 

%{\it Probability distribution:} 

%The non-diffusive and anisotropic nature of ABP dynamics are also clearly visible in position probability distribution $P(x,y,t).$  

To further characterize the anisotropic and non-diffusive early time behavior of ABP, we compute the position probability distribution function (PDF) $P(x,y,t)$. Figure \ref{fig:moments}(b) shows $P(x,y,t)$ in the $(x,y)$ plane
for $t \ll D_R^{-1}$, obtained from numerical simulations of ABP. The PDF for ABP, at early times, has an anisotropic `sickle-like' shape, with a peak near $x=v_0\,t$ and $y=0$. This is in clear contrast to the passive case where the PDF has a Gaussian (isotropic) shape with a peak at the origin.

\begin{figure}[t]
\includegraphics[width= 8.8 cm]{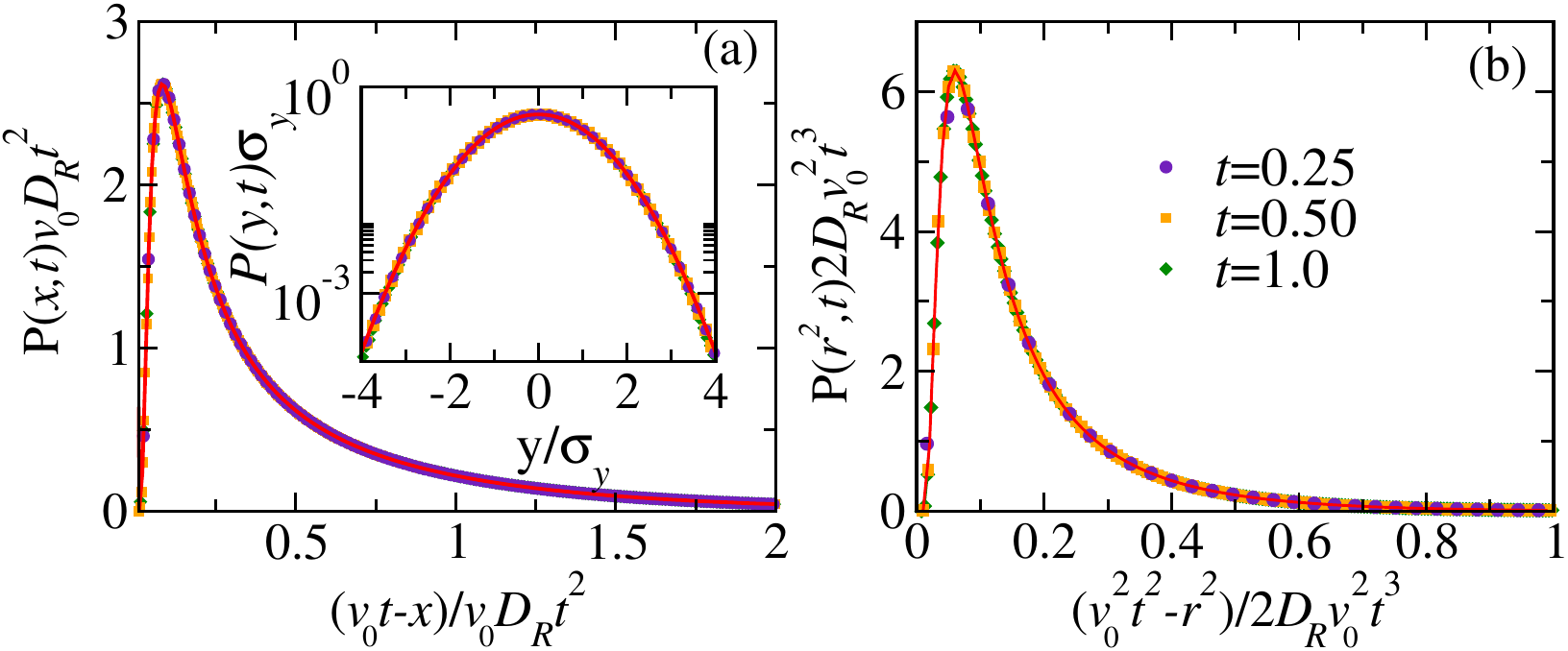} 
 \caption{Marginal distributions   $P(x,t)$ (a), $P(y,t)$ (inset of (a)) and $P(r^2,t)$ (b)  in the short-time regime for different values of $t \ll D_R^{-1}$  for $D_R=0.01$ and $v_0=1.$ The data are scaled according to the scaling forms predicted in the text and compared with the exact results computed therein. }\label{fig:Prxy}
\end{figure}

The position PDF can, in principle, be obtained by integrating over the orientational degree of freedom, i.e., $P(x,y,t) = \int_{-\infty}^{\infty} {\cal P}(x,y,\phi,t) \, d\phi$, where ${\cal P}(x,y,\phi,t)$ satisfies a Fokker-Planck (FP) equation (see Eq.~\eqref{FP_SM1} in Appendix~\ref{sec:FP}). Unfortunately, this FP equation is hard to solve at all times \cite{Mumford}. However, in the short-time regime, analytical progress can be made by using path-integral method for Brownian functionals~\cite{Satya_review}. In this limit, $\phi(t)\sim \sqrt{t}$ is small, and Eqs. (\ref{eq:modelx}) and (\ref{eq:modely}) can be approximated by $\dot x \approx  v_0 \left(1 - \frac 12 \phi^2(t) \right)$ and $\dot y \approx  v_0 \phi(t),$ keeping terms  up to ${\cal O}(\phi^2(t))$ in the expansions of $\cos \phi(t)$ and $\sin \phi(t)$.
% 
% Noting that at short-times $\phi(t)\sim \sqrt{t}$ is small, we can expand $\cos \phi(t)$ and $\sin \phi(t)$ in Eqs. (\ref{eq:modelx}) and (\ref{eq:modely}) for small $\phi(t)$. Keeping terms only up to ${\cal O}(\phi^2(t))$, we can approximate the ABP dynamics by $\dot x \approx  v_0 \left(1 - \frac 12 \phi^2(t) \right)$ and $\dot y \approx  v_0 \phi(t)$. 

It is useful to write $\phi(t)= \sqrt{2D_R} B(t)$ where 
$B(t)$ is the SBM. Using the scaling property of Brownian motion $B(\alpha t) = \sqrt{\alpha} B(t),$ the $x$ and the $y$ coordinates can be expressed~as
\bea
x(t) &=& v_0t - v_0  D_R t^2 \int_0^1 \id s~B^2(s) \cr
y(t) &=& v_0 \sqrt{2D_R} t^{3/2} \int_0^1 \id s~B(s) \;. \label{eq:xy_shortt}
\eea
The position PDF then takes the scaling form
\bea\label{joint_PDF_xy}
P(x,y,t) = \frac 1{\sqrt{2} v_0^2 D_R^{3/2}t^{7/2}} \tilde P\left(\frac{v_0t-x}{v_0 D_R t^2}, \frac{y}{v_0 \sqrt{2D_R}t^{3/2}}\right), 
\eea
where $\tilde P(a_1,a_2)$ is the joint-distribution of the two (correlated) Brownian functionals,
\bea
a_1=\int_0^1 B^2(s) \id s, \quad \text{and}\;\; a_2=\int_0^1 B(s) \id s \;. \label{eq:a1a2}
\eea
We calculate the double Laplace transform of $\tilde P(a_1,a_2)$ using path-integral approach [see Appendix~\ref{sec:a1a2} for details], and get
\bea
\left\la \exp{[-\lambda_1 a_1 - \lambda_2 a_2]} \right\ra = \frac{\exp{[\lambda_2^2~ C(\lambda_1)]}}{\sqrt{\cosh (\sqrt{2 \lambda_1})}} \label{eq:G12} 
\eea
where $C(\lambda_1) = \frac1{4\lambda_1} \left[1- \frac{\tanh(\sqrt{2 \lambda_1})}{\sqrt{2 \lambda_1}}\right]$. The scaling function  
$\tilde P(a_1,a_2)$ (and hence the PDF $P(x,y,t)$) can, in principle, be obtained by inverting the double Laplace transform in Eq.~\eqref{eq:G12}. Unfortunately, this inversion is difficult. Nevertheless, useful informations can still be extracted from this Laplace transform  as we show below. 

Integrating Eq. (\ref{joint_PDF_xy}) over $y$, the marginal distribution of $x$ clearly has the scaling form
\bea
P(x,t) = \frac 1{v_0 D_R t^2} f_x\left(\frac{v_0t-x}{v_0 D_R t^2}\right),\label{eq:Px}
\eea
where $f_x(a_1)$ is the PDF of $a_1$ defined in Eq.~\eqref{eq:a1a2}. The Laplace transform of $f_x(a_1)$ is obtained from Eq.~\eqref{eq:G12} by setting $\lambda_2 = 0,$ 
\bea
\la e^{-\lambda_1 a_1} \ra =   \int_0^\infty e^{-\lambda_1 a_1} f_x(a_1)  da_1 =  \frac{1}{\sqrt{\cosh \left(\sqrt{2\lambda_1}\right)}}.
%\left[\cosh \left(\sqrt{2\lambda_1} \right)\right]^{-1/2}.
\eea
This Laplace transform can now be explicitly inverted (see Appendix~\ref{sec:Px}) to give
\bea
f_x(a_1)= \frac 1{2 \sqrt{\pi a_1^3}} \sum_{k=0}^\infty (-1)^k \frac {(4 k+1)}{2^{2k}} \left({2k \atop k}\right) e^{-\frac{(4 k+1)^2}{8 a_1}},~~\label{eq:fx}
\eea
for $a_1\ge 0.$ The scaling function $f_x(a_1)$ is manifestly non-Gaussian with the asymptotic behaviors (see Appendix~\ref{sec:Px}):  $f_x(a_1) \simeq \frac 1{2 \sqrt{\pi a_1^3}} e^{- \frac 1{8a_1}}$ as $a_1 \to 0$ and $f_x(a_1) \simeq \frac1{\sqrt{2a_1}} e^{-\frac{\pi^2 a_1}{8}}$ as $a_1 \to \infty$. This function has a peak close to $a_1 =0$, which corresponds to $x \approx v_0\,t$ [see Fig. \ref{fig:Prxy}(a)] and it decays exponentially fast away from the edge, i.e. for $x \ll v_0 \, t$. In Fig. \ref{fig:Prxy}(a) we compare the  exact result (solid line) and the data obtained from numerical simulations for different values of $t \ll D_R^{-1}$ for a fixed $D_R.$ The excellent agreement confirms the scaling form \eqref{eq:Px} along with the analytical prediction Eq.~\eqref{eq:fx} for $t \ll D_R^{-1}$.

The marginal distribution for the $y$-component, on the other hand, takes a very different form. Taking the limit $\lambda_1 \to 0$ in Eq.~\eqref{eq:G12} we get the Laplace transform $\la \exp [-\lambda_2 a_2] \ra = \exp[\lambda_2^2/6]$. Upon Laplace inversion, $P(y,t)$ has a pure Gaussian form, $P(y,t) = \exp [- y^2/2 \sigma_y^2]/\sqrt{2 \pi \sigma_y^2}$ with the variance $\sigma_y^2=2 v_0^2 D_R t^3/3$, consistent with the first term in Eq.~(\ref{eq:xyvar}). The inset of Fig.~\ref{fig:Prxy}(a) shows $P(y,t)$ for different values of $t \ll D_R^{-1}$ as a function of $y/\sigma_y$ compared to the Gaussian (solid line), verifying the analytical~prediction.

% \bea
% P(y,t)= \sqrt{\frac 3{4 \pi v_0^2 D_R t^3}} \exp{\left[- \frac{3 y^2}{4 v_0^2 D_R t^3} \right]}.
% \eea
% Note that the variance $\sigma_y^2=2 v_0^2 D_R t^3/3$ correspond to the leading term in Eq.~\eqref{eq:xyvar}. 

For a $2d$ SBM with diffusion constant $D$, the $x$ and $y$ coordinates are completely independent (each of them is a $1d$ Brownian motion), 
and hence the position PDF in the $(xy)$ plane is simply $P(x,y,t) =  \,e^{-(x^2+y^2)/(4 D\,t)}/(4 \pi D \,t)$. Hence it is isotropic in the $(xy)$ plane and in particular the PDF of $r^2 = x^2 + y^2$ is simply $P(r^2,t) = e^{-r^2/(4D\,t)}/(4 D\,t)$. In contrast, for an ABP, the $x$ and $y$ coordinates are strongly correlated at short times and hence we expect a different behavior for $P(r^2,t)$. Indeed, this PDF $P(r^2,t)$ can also be computed explicitly by exploiting the result in Eq. (\ref{eq:G12}). From Eq.~(\ref{eq:xy_shortt}), we have $r^2=x^2 + y^2 \simeq v_0^2 t^2 - 2 D_R v_0^2 t^3 (a_1-a_2^2)$ up to order ${\cal O}(t^3)$, where $a_1$ and $a_2$ are given in Eq.~\eqref{eq:a1a2}. Thus $P(r^2,t)$, for $t \ll D_R^{-1}$, is expected to behave as
\bea\label{eq:PDFrsq}
P(r^2,t) = \frac 1{2D_R v_0^2 t^3} f_r \left(\frac{v_0^2t^2-r^2}{2 D_R v_0^2 t^3} \right)
\eea
where $f_r(z)$ is the probability distribution of $z=a_1-a_2^2$. Its Laplace transform can be extracted from Eq.~\eqref{eq:G12} after a few steps of algebra detailed in Appendix~\ref{sec:Pr2},
\bea
\la e^{-\lambda z} \ra= \int_0^\infty e^{-\lambda z} \, f_r(z)\, dz = \left[\frac{\sqrt{2 \lambda}}{\sinh (\sqrt{2 \lambda})}\right]^{1/2} \;,
\eea
which, fortunately, can be inverted explicitly. The resulting expression, given in Eq. \eqref{LTinv6} in Appendix~\ref{sec:Pr2}, is somewhat long but explicit. 
It has the asymptotic behaviors: $f_r(z) \simeq \frac 1{2 \sqrt{2 \pi}z^2} e^{-1/(8z)}$ as $z \to 0$ and $f_r(z) \simeq \sqrt{\frac \pi z} e^{-\pi^2 z/2}$ as $z \to \infty$. This scaling function $f_r(z)$ is plotted and compared to simulations in Fig.~\ref{fig:Prxy}(b) with excellent agreement. 

Let us remark that this radial distribution $P(r^2,t)$ in Eq. (\ref{eq:PDFrsq}), being an average over the angular degree, does not provide any information about the anisotropy present in the $(xy)$ plane. It just demonstrates that the strong correlations between the $x$ and $y$ coordinates at short times make the radial distribution very different from that of the SBM case. Even if one averages over the initial orientation angle $\phi(0)$ uniformly over $[0,2 \pi]$, thus restoring isotropy in the $(xy)$ plane at all times, $P(r^2,t)$ is still given by the same expression as in (\ref{eq:PDFrsq}), which manifestly is still different from the SBM. This rotationally symmetric case was qualitatively discussed in Ref.~\cite{Sevilla, Sevilla2}, although no analytical form for the distribution was found. Very recently, the Fourier transform of the PDF was computed as a formal eigenfunctions expansion in terms of the Mathieu functions \cite{kurz}. However, inverting this formal Fourier transform and plotting it in real space is still highly difficult. Our approach provides an explicit real space distribution, which is exact at short times.

%\vspace*{0.2cm}
%\noindent{\it First-passage properties:} 

\section{First-passage properties}

\begin{figure}[t]
 \centering
 \includegraphics[width=8.8cm]{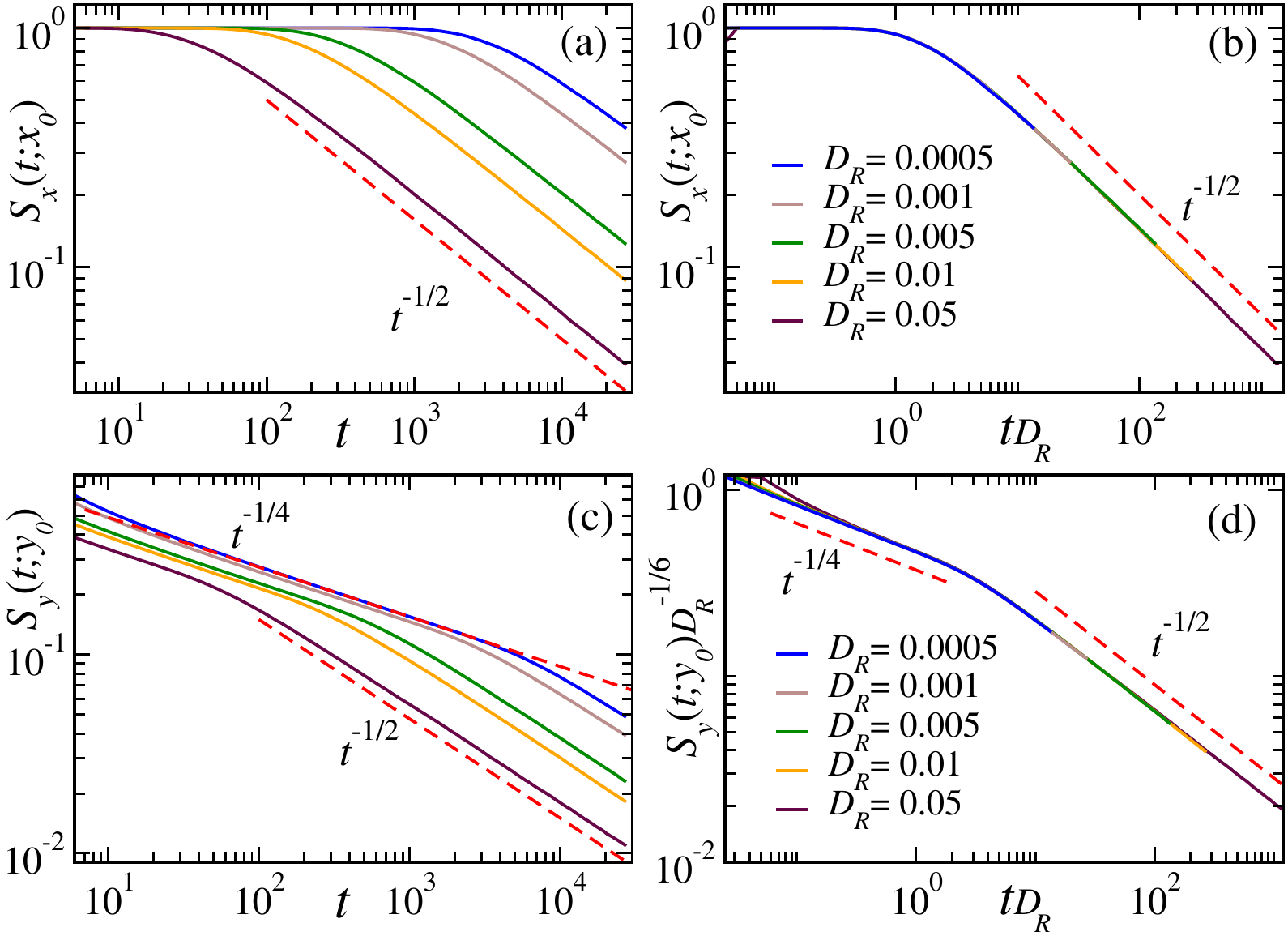}
 % Syx.pdf: 770x289 pixel, 72dpi, 27.16x10.20 cm, bb=0 0 770 289
 \caption{Survival probability: (a) Plot of $S_x(t;x_0)$ as a function of time $t$ for different values of $D_R$ and the initial position $x_0=0.1, y_0=0.$ (b) Collapse of the curves in (a) following Eq.~\eqref{eq:Sx_scaling}. (c) $S_y(t;y_0)$ $vs$ $t$ for different values of $D_R$ and the initial position  $x_0=0, y_0=0.1.$ The curves
  initially show a power law decay $t^{-1/4}$ which crosses over to $t^{-1/2}$ after a time $t \sim D_R^{-1}$. The topmost dashed line is the exact prediction from the RAP in Eq.~\eqref{eq:Sy_RAP}. (d)  Collapse of the curves in (c) according to Eq.~\eqref{eq:Sy_scaling}. Here $v_0=1$ for all the curves.}
 \label{fig:Syx}
\end{figure}

The discussion above clearly shows a crossover from early time, non-diffusive and anisotropic behavior (a fingerprint of `activeness') to the late time diffusive and isotropic Brownian behavior, at a crossover time $t \sim D_R^{-1}$. Another natural observable that demonstrates this crossover in a candid way is the first-passage probability. In fact, for active systems, the first-passage properties have not been explored much, except very recently in a class of one-dimensional models ~\cite{Angelani,Dhar2017,Scacchi,Maes18}. The first-passage probability is most conveniently defined through the survival probability~\cite{satya_review,Bray,Redner}. Let us start with the $x$-coordinate and  
denote by $S_x(t;x_0)$ the probability that the $x$-component, starting at $x_0\geq 0$, does not cross $x=0$ up to time $t$ (clearly it does not depend on $y$). The associated first-passage probability is just $-\partial_t \,S_x(t;x_0)$. In this case, at early times $t \ll D_R^{-1}$, $x(t) \approx v_0\,t$ with $v_0>0$, as in Eq. (\ref{eq:xy_shortt}), and therefore $x(t)$ stays positive with probability close to unity. However, at long times $t \gg D_R^{-1}$, $x(t)$ behaves diffusively and we would expect~\cite{satya_review,Bray,Redner} a decay $\sim t^{-1/2}$ of $S_x(t;x_0)$ at late times. For simplicity, we set $x_0\to 0$ and in this case, we naturally expect a crossover behavior of the form 
\bea 
S_x(t;x_0 \to 0) = {\cal F}_{x}\left( t \, D_R \right) \;, \label{eq:Sx_scaling}
\eea
where the crossover function ${\cal F}_{x}(u) \sim 1$ for $u \ll 1$, while  ${\cal F}_x(u) \sim u^{-1/2}$ for $u \gg 1$. Figures \ref{fig:Syx}(a) and (b) show the behavior of $S_x(t;x_0 \to 0)$  for different values of $D_R$. The collapsed data in Fig. \ref{fig:Syx}(b) are in agreement with the proposed scaling form in Eq.~(\ref{eq:Sx_scaling}).

We expect the first-passage probability for the $y$-coordinate to be rather different due to the anisotropy present
in early times. As in the $x$-case, we denote by $S_y(t;y_0)$ the survival probability that the $y$-coordinate, starting initially at $y_0 > 0$, stays positive up to time $t$. At times $t \ll D_R^{-1}$, the effective Langevin equation for the $y$-coordinate, $\dot y \approx v_0 \phi(t)$, can be recast (after taking one more time-derivative) as    
\bea
\ddot y \approx \sqrt{2\, D_{\textrm{RAP}}} \;\,\eta_{\phi}(t) \;\;, \;\; {\textrm{where}} \;\;  D_\text{RAP}= v_0^2 D_R \;. \label{eq:RAM}
\eea
This effective Langevin equation for the $y$-coordinate is thus exactly identical to the celebrated `Random Acceleration Process' (RAP), which has been studied extensively in the literature as one of the simplest non-Markovian processes~(see \cite{Burkhardt2007,satya_review,Bray,Burkhardtbook,Masoliver}). One hallmark of this non-Markovian nature is an anomalously slow decay of the survival probability $S_y(t;y_0) \sim t^{-1/4}$ (for $t \gg 1$)~\cite{Burkhardt, Burkhardt2007, satya_review,Bray,Burkhardtbook}, in marked contrast to the $t^{-1/2}$ decay of the SBM~\cite{satya_review,Bray,Redner}. In the limit $y_0 \to 0$, the full survival probability (and not just the persistence exponent $1/4$) can be computed~\cite{RAP}. Translating this result in our units we then predict, for $t \gg 1$ 
\bea
S_y(t;y_0) \simeq \frac {2^{ 5/6}\Gamma(-4/3)}{3^{2/3}\pi \Gamma(3/4)} \left(\frac{y_0\, D_R }{v_0}\right)^{1/6} \left(t\,D_R \right)^{-1/4}  \;.\label{eq:Sy_RAP}
\eea
Since the mapping to the RAP holds only for $t \ll D_R^{-1}$, we expect this result in (\ref{eq:Sy_RAP}) to hold for $1 \ll t \ll D_R^{-1}$. 
For $t \gg D_R^{-1}$, given that the ABP behaves as a SBM, we would expect $S_y(t;y_0)$ to decay as $t^{-1/2}$. This suggests a crossover from the early time $\sim t^{-1/4}$ to the late time $t^{-1/2}$ decay of $S_y(t;y_0)$, described by the scaling form  
\bea
S_{y}(t;y_0) = \left(\frac{y_0 \, D_R}{v_0}\right)^{1/6} {\cal F}_{y}\left(t\, D_R \right) \;, \label{eq:Sy_scaling}
\eea
where the crossover scaling function ${\cal F}_{y}(u)$ has the limiting behaviors: ${\cal F}_{y}(u) \sim u^{-1/4}$ for $u \ll 1$ and ${\cal F}_{y}(u) \sim u^{-1/2}$ for $u \gg 1$. 
%
%\bea
%{\cal F}_{y}(u) \sim \left \{ \begin{split}                             
%u^{-1/4} \quad \text{for}\;\; u \ll 1 \cr
%u^{-1/2} \quad \text{for}\;\; u \gg 1. 
%                              \end{split}
%                              \right. \label{eq:Fy}
%\eea
%
We have verified this scaling form (\ref{eq:Sy_scaling}) numerically. Figure \ref{fig:Syx}(c) shows $S_y(t;y_0)$ vs. $t,$ obtained from simulations, for different values of $D_R.$ The uppermost dashed line in Fig.~\ref{fig:Syx}(c) is the exact prediction from Eq.~\eqref{eq:Sy_RAP}. The same data scaled according to Eq.~\eqref{eq:Sy_scaling} are plotted in Fig \ref{fig:Syx}(d). The excellent data collapse confirms the predicted scaling form \eqref{eq:Sy_scaling}.

%{\it Conclusion:} 
\section{Conclusion}

We have studied the dynamics of an ABP in $2d$. The rotational diffusion sets a timescale $D_R^{-1}$ for the dynamics of the spatial coordinates of the ABP. While for $t \gg D_R^{-1}$ the ABP reduces to an ordinary $2d$ Brownian motion, we have shown that the fingerprint of the `activeness' of motion shows up  
for $t \ll D_R^{-1}$, where the dynamics is highly anisotropic with strong correlations between the $x$ and the $y$-coordinates. For short times, we have computed exactly the marginal distributions $P(x,t)$, $P(y,t)$ as well as the radial marginal distribution $P(r^2,t)$, well verified by numerical simulations. In addition, we have shown that, at these early times, the ABP has anomalous first-passage properties. In particular, the survival probability in the $y$-direction decays anomalously as $t^{-1/4}$ for $1 \ll t \ll D_R^{-1}$.

An alternative way to explore the anomalous `active' behavior is to apply an external confining potential such as a harmonic trap $V(r) = \mu\,r^2/2$, that introduces an additional time scale $\mu^{-1}$. By tuning $\mu$, one can show that the anomalous early time fingerprints of the free ABP studied here can be made to persist even at late times, leading to non-Boltzmann stationary states~\cite{us_tocome}. Indeed, a crossover from Boltzmann (passive) to non-Boltzmann (active) stationary states in the presence of a trap 
has been studied using numerical simulations and scaling arguments~\cite{Solon2015, Potosky2012}, as well as in recent experiments on trapped self-propelled particles~\cite{Takatori,dauchot}. By tuning these two timescales, $D_R^{-1}$ and $\mu^{-1}$, it would be interesting to see if our analytical predictions at short times for the free ABP can be observed experimentally.

%\acknowledgments

\begin{acknowledgments}
 
The authors would like to acknowledge the support from the Indo-French Centre for the promotion of advanced research (IFCPAR) under Project No. 5604-2. We thank I. Dornic and J. M. Luck for useful discussions and for pointing out Ref. \cite{Mumford}. We would also like to thank ICERM for hospitality, where part of this work has been done. 
%This work was partially supported by ANR grant ANR-17-CE30-0027-01 RaMaTraF.
 
\end{acknowledgments}

\appendix

%\noindent{\bf Effective Noise:} 

\section{Effective Noise}\label{sec:noise}

The Langevin equation \eqref{eq:model} %in the main text 
reads
\bea\label{langevin_sm}
\dot x &=& v_0 \cos \phi(t) = \xi_x(t) \cr
\dot y &=& v_0 \sin \phi(t)  = \xi_y(t)
\eea
where $\xi_x(t)$ and $\xi_y(t)$ are the effective noises in the $x$ and the $y$ direction respectively. Here, $\phi(t)$ is a Brownian motion with a diffusion constant $D_R$ [see Eq.~(1) in the main text] starting from $\phi(0)=0$. Clearly, $\phi(t)$ is a Gaussian process with two-point correlation
\bea\label{eq:correl}
\langle \phi(t_1) \phi(t_2)\rangle = 2 \, D_R \min(t_1,t_2) \;.
\eea 
The noise $\xi_{x,y}(t)$ are bounded in time, even though $\phi(t)$ grows with time as $\phi(t) \sim \sqrt{2\,D_R \, t}$.
It is convenient to use the exponential forms for the noise, $\xi_x(t) = \frac 12 \left(e^{i \phi(t)}+e^{-i \phi(t)} \right)$ and $\xi_y(t) = \frac 1{2i} \left(e^{i \phi(t)}-e^{-i \phi(t)}\right).$ The one and two-point correlation functions of the noise $\xi_{x,y}(t)$ can then be computed using the fact that $\phi(t)$ is a Gaussian process. In fact, we will use a well known identity for a Gaussian process $\phi(t)$,
\begin{equation}\label{identity_Gauss}
\Big \langle \exp{\big(i \sum_j b_j \, \phi(t_j)\big)}  \Big\rangle = \exp{\Big(-\frac{1}{2} \sum_{j,k} b_j b_k \la \phi(t_j) \phi(t_k)\ra \Big)}
\end{equation}
where $b_i$'s are arbitrary. The average values are thus given by  
\bea
\la \xi_x(t) \ra  = v_0 e^{-D_R t}\;, \la \xi_y(t) \ra  = 0 \label{eq:xi_av}
\eea
where we used Eq. (\ref{identity_Gauss}) with appropriate values of $b_1$ (and $t_1 = t$) and $b_j=0$ for $j > 1$ and $\la \phi^2(t)\ra = 2 D_R\, t$. Similarly the two-point functions can be obtained in a straightforward manner from Eq. (\ref{identity_Gauss}) by appropriately choosing $b_1$ and $b_2$ and keeping $b_j = 0$ for $j>2$. We get
\bea%\label{twopointcorrel}
\la \xi_x(t_1) \xi_x(t_2)\ra &=& \frac {v_0^2}{2} \bigg[ e^{-D_R |t_1-t_2|} + e^{-D_R\left(t_1 + t_2 + 2 \min(t_1,t_2)\right)} \bigg]. \cr
&& \label{eq:xixx}
\eea
In the limit of $t_1 \to \infty, t_2 \to \infty$ but finite $|t_1 - t_2|$ the above expression reduces to Eq.~(2) of the main text. The-autocorrelation of $\xi_y(t)$ can also be computed similarly, yielding
\bea
\la \xi_y(t_1) \xi_y(t_2)\ra &=& \frac {v_0^2}{2}  \bigg[ e^{-D_R |t_1-t_2|} - e^{-D_R\left(t_1 + t_2 + 2 \min(t_1,t_2)\right)}\bigg] \cr
&& \label{eq:xiyy}
\eea
which has the same limiting behavior as the correlator of $\xi_x$ for large $t_1, t_2$ (see Eq. (2) in the main text). The mutual two-point correlation $\la \xi_x(t_1) \xi_y(t_2)\ra$ vanishes due to symmetry, however, higher order mutual correlations remain finite. This is easily seen from the exact relation $\xi_x^2(t) + \xi_y^2(t) = v_0^2$, using Eq. (\ref{langevin_sm}). 
%
% 
% \bea\label{twopointcorrel}
% &&\la \xi(t_1) \xi(t_2) \ra =\la \bar \xi(t_1) \bar \xi(t_2) \ra \\
% &&=v_0^2 \exp{\left[-D_R\left(t_1 + t_2 + 2 \min(t_1,t_2)\right) \right]} 
% \eea
% where $\bar \xi(t) = v_0 e^{-i\phi(t)}$ is the complex conjugate of $\xi(t)$. Similarly we also obtain
% \bea
% &&\la \xi(t_1) \bar \xi(t_2) \ra = \la \bar \xi(t_1) \xi(t_2) \ra \\
% &&=v_0^2 \exp [-D_R |t_1-t_2|] \;. \label{eq:xi_xibar}
% \eea
% This result is quoted as Eq.~(3) in the main text. 
%
%:
From Eqs. (\ref{eq:xixx}) and (\ref{eq:xiyy}) one sees that the effective noises $\xi_{x,y}(t)$ appearing in 
the Langevin equation (\ref{langevin_sm}), have a finite correlation time $D_R^{-1}$, indicating that they have a finite memory.

%The stationary marginal distribution of the components $\xi_x= v_0 \cos \phi$ and $\xi_y =v_0 \sin \phi$ of the effective noise can also be obtained. \\
  
%\vspace*{0.3cm}  
 % \noindent{\bf Fokker-Planck Equation:} 
\section{Fokker-Planck Equation}\label{sec:FP}
 
Let ${\cal P}(x,y,\phi,t)$ denote the probability that at any time $t,$ the ABP has a position $(x,y)$ and orientation $\phi.$ ${\cal P}(x,y,\phi,t)$ evolves according to a Fokker-Planck equation,
\bea\label{FP_SM1}
\partial_t {\cal P}(x,y,\phi;t) &=& - v_0 \left[ \cos \phi\frac{\partial {\cal P}}{\partial x}
+ \sin \phi \frac{\partial {\cal P} }{\partial y} \right]  + D_R \frac{\partial^2 {\cal P}}{\partial \phi^2} \cr
&& 
\eea
where we have supressed the argument of ${\cal P}$ on the right hand side for brevity. The marginal probability distribution 
of the position can then be obtained by integrating over $\phi,$
\bea
P(x,y,t) = \int_{-\infty}^{\infty} \id \phi ~{\cal P}(x,y,\phi,t) \;.
\eea
% In the long time limit the position distribution $P(x,y,t)$ converges to a stationary form  which is denoted by 
% \bea\label{Pstat}
% P_{\textrm{stat}}(x,y) = P(x,y,t \to \infty) \;.
% \eea

%\vspace*{0.3cm}
  %\noindent{\bf Mean-squared displacements:} 

\section{Mean-squared displacements}\label{sec:MSD}  
  
The mean-squared displacement (MSD) of the ABP can be exactly calculated from the Langevin equation (\ref{langevin_sm}). %, upon using the noise correlations in Eqs.~\eqref{eq:xi_av}, \eqref{eq:xixx} and \eqref{eq:xiyy}. 
The average displacements along the $x$ and $y$ directions immediately follow from \eqref{eq:xi_av}
\bea \label{xy_av}
\la x(t) \ra &=& \int_0^t \id s \la \xi_x(s)\ra = \frac{v_0}{D_R} \left(1 - e^{-D_R t} \right) \\
\la y(t) \ra &=& \int_0^t \id s \la \xi_y(s)\ra = 0 \;.
\eea
The computation of second moments involves the noise-correlators given in Eqs.~\eqref{eq:xixx} and \eqref{eq:xiyy},
\bea
\la x^2(t) \ra &=& \int_0^t ds_1 \, \int_0^t ds_2 \, \la \xi_x(s_1) \xi_x(s_2)\ra \cr
\la y^2(t) \ra &=& \int_0^t ds_1 \, \int_0^t ds_2 \, \la \xi_y(s_1) \xi_y(s_2)\ra. 
\eea
Evaluating the above integrals, we get the variances of the $x$ and the $y$-components separately,
\bea\label{xy_var_expl}
\sigma_x^2 &=& \la x^2 \ra - \la x \ra^2 \cr
&=& \frac{v_0^2}{D_R} t +\frac{v_0^2}{12 D_R^2} \left[e^{-4 D_R t} - 12e^{- 2D_R t} +32 e^{- D_R t} -21 \right] \cr
\sigma_y^2 &=& \la y^2 \ra  \cr
&=& \frac{v_0^2}{D_R}t  - \frac{v_0^2}{12 D_R^2}\left[e^{-4 D_R t} - 16 e^{- D_R t}+15 \right] \;.
\eea
At long times $t \gg D_R^{-1}$, both $\sigma_x^2 \approx 2\,D_\text{eff} \,t$ and $\sigma_y^2 \approx 2\,D_\text{eff} \,t$ grow linearly with time with an effective diffusion constant $D_\text{eff} = v_0^2 /(2\,D_R)$. In contrast, at short-times $t \ll D_R^{-1}$, expanding (\ref{xy_var_expl}) in Taylor series, we find 
\bea
\sigma_x^2 &\approx& \frac{1}3 v_0^2 D_R^2 t^4  - \frac 7{15} v_0^2 D_R^3 t^5 + \cdots \cr
\sigma_y^2 &\approx& \frac{2}3 v_0^2 D_Rt^3  - \frac 5{6} v_0^2 D_R^2 t^4 + \cdots.
\eea
which gives the results in Eqs.~(3) and (4) of the main text. These results reflect a strong anisotropy at early times in the $(x,y)$ plane since, for small $t$, $\sigma_y^2 \sim t^3 \gg \sigma_x^2 \sim t^4$. 

Note that, from these results above, we can also compute the radial mean-squared displacement, defined as
\bea\label{rsq_1}
\la r^2(t) \ra = \la x^2(t)\ra + \la y^2(t)\ra = \sigma_x^2 + \la x \ra^2 + \sigma_y^2 \;. 
\eea
Using the results from Eqs. (\ref{xy_av}) and (\ref{xy_var_expl}) this gives
\bea\label{rsq_2}
\la r^2(t) \ra = \frac{2v_0^2}{D_R} t  + \frac{2v_0^2}{D_R^2} \left(e^{-D_R t}-1 \right) \;.
\eea
This result for the radial MSD was in fact known for a long time~\cite{BechingerRev}. However, the individual variances along the $x$ and the $y$ directions in Eq. (\ref{xy_var_expl}), which clearly illustrate the anisotropy, have not been computed so far, to the best of our knowledge.

%\vspace*{0.3cm}
%\noindent {\bf Joint distribution of $a_1$ and $a_2$:} 

\section{Joint distribution of $a_1$ and $a_2$}\label{sec:a1a2}

Let $\tilde P(a_1,a_2)$ denote the joint probability distribution of the Brownian functionals
\bea
a_1=\int_0^1 B^2(s) \id s, \quad \text{and}\;\; a_2=\int_0^1 B(s) \id s \;. \label{eq:Sa1a2}
\eea
where $B(t)$ is the standard Brownian motion $B(t),$ satisfying $\dot B(t) = \eta(t)$, where $\eta(t)$ is a delta-correlated white noise of zero mean. We calculate the double Laplace transform,
\bea
Q(\lambda_1,\lambda_2) &=& \left\la \exp{[-\lambda_1 a_1 - \lambda_2 a_2]} \right\ra \cr
&=& \int_0^{\infty} \id a_1\int_{-\infty}^{\infty}\id a_2 ~ e^{-\lambda_1 a_1 - \lambda_2 a_2} \tilde P(a_1,a_2)\cr
&&
\eea
Using the Brownian path measure, $Q(\lambda_1,\lambda_2)$ can be expressed as a path integral,

\begin{widetext}
\bea \label{Q1}
 Q(\lambda_1,\lambda_2) &=& \int_{-\infty}^{\infty} \id x \int_{B(0)=0}^{B(1)=x} {\cal D}B(\tau) \exp{\left[-\frac 12 \int_0^1 \id \tau \dot B^2(\tau)  - \lambda_1 \int_0^1 \id \tau B^2(\tau)  -\lambda_2 \int_0^1 \id \tau~B(\tau) \right]} \;,
%&=& \int_{-\infty}^{\infty} \id x \int_{B(0)=0}^{B(1)=x} {\cal D}B(\tau) \exp{\left[-\frac 12 \int_0^1 \id \tau \dot B^2(\tau)  - \lambda_1 \int_0^1 \id \tau \left\{\left(B(\tau) + \frac{\lambda_2}{2 \lambda_1}\right)^2 - \frac{\lambda_2^2}{4 \lambda_1^2} \right\} \right]} \cr
%&=& e^{\frac{\lambda_2^2}{4 \lambda_1}}\int_{-\infty}^{\infty} \id x ~{\cal G}\left(x,\frac{\lambda_1}{2 \lambda_2},1 \right)
\eea
\end{widetext}
where $B(\tau)$ is a Brownian motion that starts at $B(0) = 0$ at time $\tau=0$ and arrives at a final position $B(1) = x$ at time $\tau = 1$. One then integrates over the final position $x$. To evaluate this path integral, we complete the square and rewrite it as follows
\begin{widetext}
 \bea \label{Q2}
 Q(\lambda_1,\lambda_2) =  \int_{-\infty}^{\infty} \id x \int_{B(0)=0}^{B(1)=x} {\cal D}B(\tau) \exp{\left[-\frac 12 \int_0^1 \id \tau \dot B^2(\tau)  - \lambda_1 \int_0^1 \id \tau \left\{\left(B(\tau) + \frac{\lambda_2}{2 \lambda_1}\right)^2 - \frac{\lambda_2^2}{4 \lambda_1^2} \right\} \right]}  \;.
\eea
\end{widetext}
Next, we make a shift $\tilde B(\tau) = B(\tau) + \lambda_2/(2 \lambda_1)$ and also change $x \to x + \lambda_2/(2 \lambda_1)$. This constant shift does not change the Brownian measure, since $\dot{\tilde B}(\tau) = \dot B(\tau)$. Hence $\tilde B(\tau)$ is another Brownian motion that starts at $\tilde B(0) = \lambda_2/(2 \lambda_1)$ and arrives at $x$ at $\tau = 1$. For simplicity of notation, we will re-denote $\tilde B(\tau)$ as $B(\tau)$ with $B(0) = \lambda_2/(2 \lambda_1)$ and $B(1) = x$. Therefore Eq. (\ref{Q2}) can be written as
\begin{widetext}
\bea \label{Q3}
 Q(\lambda_1,\lambda_2) &=& e^{\frac{\lambda_2^2}{4 \lambda_1}}\int_{-\infty}^{\infty} \id x \int_{B(0)=\lambda_2/(2 \lambda_1)}^{B(1)=x} {\cal D}B(\tau) \exp{\left[-\frac 12 \int_0^1 \id \tau \dot{B}^2(\tau)  - \lambda_1 \int_0^1 \id \tau B^2(\tau)   \right]} \;.
 \eea
 \end{widetext}
 The form of the path integral in Eq. (\ref{Q3}) shows immediately that this corresponds to the imaginary time propagator of a quantum harmonic oscillator with Hamiltonian  $\hat H = -\frac{\hbar^2}{2m} \frac{\partial^2}{\partial x^2}+ \frac12 m\, \omega^2 x^2$, upon setting $\hbar = m =1$ and $\omega = \sqrt{2 \lambda_1}$. It propagates from the initial position $\lambda_2/(2 \lambda_1)$ to the final position $x$ in unit time. Hence we can write 
\bea \label{Q4}
 Q(\lambda_1,\lambda_2) &=&   e^{\frac{\lambda_2^2}{4 \lambda_1}}\int_{-\infty}^{\infty} \id x ~{\cal G}\left(x,\frac{\lambda_2}{2 \lambda_1},1 \right) \;,
\eea 
where ${\cal G}(x,y,t)$ denotes the imaginary time propagator of the quantum harmonic oscillator from the initial position $y$ to the final position $x$ in imaginary time $t$.  This propagator is well-known in the literature \cite{Feynman} and for a given $\omega$, with $\hbar = m = 1$ it reads
\begin{widetext}
\bea
{\cal G}(x,y,t)= \sqrt{\frac{\omega}{2 \pi \sinh (\omega\,t)}} \exp\left[- \frac{\omega}{2 \sinh (\omega\, t)}[(x^2+y^2)\cosh(\omega\,t) - 2 xy]\right] \;.
 \label{eq:Gxy1}
\eea
 \end{widetext}
In our case, setting $\omega= \sqrt{2\lambda_1}$ and $y=\lambda_2/2 \lambda_1$, and performing the Gaussian integration over $x$ we finally get
\bea
Q(\lambda_1, \lambda_2) = \left\la \exp{[-\lambda_1 a_1 - \lambda_2 a_2]} \right\ra = \frac{\exp{[\lambda_2^2~ C(\lambda_1)]}}{\sqrt{\cosh (\sqrt{2 \lambda_1})}},\;\;\; \label{eq:G12A} 
\eea
where
\bea \label{eq:C}
C(\lambda_1) = \frac{1}{4\lambda_1} \left(1 - \frac{\tanh{\sqrt{2\lambda_1}}}{\sqrt{2 \lambda_1}} \right) \;.
\eea
This then provides the derivation of Eq.~\eqref{eq:G12} in the main text. Formally, the joint distribution $\tilde P(a_1,a_2)$ can be obtained by inverting the double Laplace transform,
\bea\label{inverse_a1a2}
\tilde P(a_1,a_2) = \int \frac{\id \lambda_1}{2 \pi i} \int \frac{\id \lambda_2}{2 \pi i}~ Q(\lambda_1,\lambda_2) \exp{[\lambda_1 a_1 + \lambda_2 a_2]}\n
\eea
where both integrals are understood as Bromwich integrals for complex $\lambda_1$ and $\lambda_2$.

%\vspace*{0.3cm}
%\noindent {\bf Marginal distribution of $x(t)$:} 

\section{Marginal distribution of $x(t)$}\label{sec:Px}

From Eq. \eqref{eq:xy_shortt}, we have for time $t \ll D_R^{-1}$
\bea\label{x_of_t}
x(t) = v_0 \,t - v_0 D_R\, t^2\, a_1 \;,
\eea 
where $a_1 = \int_0^1 ds\, B^2(s)$. Eq. \eqref{eq:Px}  then immediately follows from this relation where $f_x(a_1)$ is the PDF of $a_1$. Setting $\lambda_2 = 0$ in Eq. (\ref{eq:G12A}), we get
\bea\label{LTa1}
\langle e^{-\lambda_1 a_1}\rangle = \int_0^\infty e^{-\lambda_1 a_1} f_x(a_1)\, da_1 = \frac{1}{\sqrt{\cosh{(\sqrt{2 \lambda_1})}}} \;.
\eea
To invert this Laplace transform, we first re-write the right hand side (r.h.s.) as
\bea\label{LTa1_2}
\frac{1}{\sqrt{\cosh{(\sqrt{2 \lambda_1})}}} = \frac{\sqrt{2} e^{-\sqrt{\frac{\lambda_1}{2}}} }{(1 + e^{-2 \sqrt{2 \lambda_1}})^{1/2}} \;.
\eea
Performing a formal series expansion of Eq. (\ref{LTa1_2}) we get
\bea\label{LTa1_3}
\langle e^{-\lambda_1 a_1}\rangle = \sqrt{2} \sum_{k=0}^\infty (-1)^k \frac{(2k)!}{(k!)^2 2^{2k}} e^{-\frac{1}{\sqrt{2}}(1+4 k) \sqrt{\lambda_1}} \;.
\eea
We can now invert term by term using the Laplace inversion identity (for $b>0$)
\bea\label{laplace_inv1}
{\cal L}^{-1}_{s \to x} \left[e^{-b \sqrt{s}}\right] = \frac{b}{2 \sqrt{\pi x^3}} e^{-\frac{b^2}{4x}} \;.
\eea  
This then gives
\bea\label{LTa1_4}
f_x(a_1)= \frac 1{2 \sqrt{\pi a_1^3}} \sum_{k=0}^\infty (-1)^k \frac {(4 k+1)}{2^{2k}} \left({2k \atop k}\right) e^{-\frac{(4 k+1)^2}{8 a_1}}, \;\; \;
\eea
stated as the formula \eqref{eq:fx} in the main text. The asymptotic behaviour of $f_x(a_1)$ in the limit $a_1 \to 0$ is simple to obtain. Indeed, for small $a_1$, the series in (\ref{LTa1_4}) is dominated by the $k=0$ term and we get 
\bea\label{a1_small}
f_x(a_1) \simeq \frac{1}{2\sqrt{\pi a_1^3}}\, e^{-\frac{1}{8 a_1}} \;.
\eea
Thus the function has an essential singularity as $a_1 \to 0$. To derive the asymptotic behavior as $a_1 \to \infty$ turns out to be trickier, as this series representation in Eq. (\ref{LTa1_4}) is not convenient to analyse in that limit. We therefore need a different representation that is well suited for the $a_1 \to \infty$ limit. To proceed, we use the following identity
\bea\label{id_cosh}
\frac{1}{\cosh(\sqrt{2 \lambda})} = 2 \sum_{n=0}^\infty (-1)^n \frac{(n+\frac{1}{2})\pi}{(n+\frac{1}{2})^2 \pi^2 + 2 \lambda} \;,
\eea
which can be simply obtained by computing the residues at the zeroes of $\cosh{(\sqrt{2 \lambda})}$ in the complex $\lambda$ plane. Therefore, we can re-write Eq. (\ref{LTa1}) as
\bea\label{id_cosh2}
\int_0^\infty e^{-\lambda_1 a_1} f_x(a_1)\, da_1 = \left[ 2 \sum_{n=0}^\infty  \frac{(-1)^n(n+\frac{1}{2})\pi}{(n+\frac{1}{2})^2 \pi^2 + 2 \lambda}\right]^{1/2}.\;\;\;
\eea
For large $a_1$, the dominant contribution comes from the branch-cut around the pole corresponding to the $n=0$ term in this series (\ref{id_cosh2}). Near this pole, the r.h.s. behaves as $\sqrt{\pi/(\pi^2/4 + 2 \lambda)}$. This can be inverted using the identity
\bea\label{id_cosh3}
{\cal L}^{-1}_{\lambda \to x} \left[\frac{1}{\sqrt{\lambda+b}}\right] = \frac{1}{\sqrt{\pi x}} e^{-b\,x} \;.
\eea
This gives
\bea\label{a1_large}
f_x(a_1) \simeq \frac{1}{\sqrt{2a_1}} e^{- \frac{\pi^2}{8}\,a_1} \;,
\eea
as stated in the main text.

\vspace*{0.3cm}
%\noindent {\bf Marginal distribution of $y(t)$:} 
\section{Marginal distribution of $y(t)$}\label{sec:Py}

From Eq. \eqref{eq:xy_shortt}, we see that
\bea\label{y_of_t}
y(t) = v_0 \sqrt{2\,D_R}\,t^{3/2} a_2 \;,
\eea
where $a_2 = \int_0^1 ds B(s)$. This immediately shows that for $t \ll D_R^{-1}$
\bea\label{Pofy1}
P(y,t)  = \frac{1}{v_0\sqrt{2 D_R}\, t^{3/2}} f_y\left( \frac{y}{v_0\sqrt{2 D_R}\, t^{3/2}}\right) \;,
\eea
where $f_y(a_2)$ is the PDF of $a_2 = \int_0^1 ds\, B(s)$. The Laplace transform of $f_y(a_2)$ can be obtained by setting $\lambda_1 = 0$ in Eq. (\ref{eq:G12A}). In the limit $\lambda_1 \to 0$, $C(\lambda_1) \to 1/6$ in Eq. (\ref{eq:C}). Hence we get
\bea\label{LTa2}
\langle e^{-\lambda_2 a_2}\rangle = \int_{-\infty}^\infty e^{-\lambda_2 a_2} f_y(a_2)\, da_2 =\exp\left(\frac{\lambda_2^2}6\right) \;.
\eea
This can be easily inverted to yield a purely Gaussian scaling function 
\bea\label{LTa2_2}
f_y(a_2)  = \sqrt{\frac{3}{2 \pi}}\exp\left(- \frac{3}{2} \, a_2^2 \right) \;.
\eea
This result is stated in the main text, in a slightly different but equivalent form.

%\vspace*{0.3cm}
%\noindent {\bf Marginal distribution of $r^2(t) = x^2(t) + y^2(t)$:} 

\section{Marginal distribution of $r^2(t) = x^2(t) + y^2(t)$}\label{sec:Pr2}

From Eq. \eqref{eq:xy_shortt}, we obtain, to leading order for small time $t \ll D_R^{-1}$
\bea \label{rsq}
r^2(t) = x^2(t) + y^2(t) = v_0^2 t^2 - 2 v_0^2 D_R t^3 \left(a_1 - a_2^2 \right) \;,
\eea
where $a_1 = \int_0^1 B^2(s) ds$ and $a_2 = \int_0^1 B(s) ds$. This immediately yields the scaling form given in Eq. (12) of the main text where $f_r(z)$ is the PDF of $z = a_1 - a_2^2$. By definition, the distribution of $z$ is thus given by
\bea\label{PDFr_1}
f_r(z) = \int_0^\infty da_1 \int_{-\infty}^\infty da_2 \,\tilde P(a_1,a_2) \delta(z-(a_1-a_2^2)),\;\;
\eea
in terms of the joint PDF of $a_1$ and $a_2$ given in Eqs. (\ref{inverse_a1a2}) and (\ref{eq:G12}). From these equations (\ref{inverse_a1a2}) and (\ref{eq:G12A}), we notice that the inverse Laplace transform with respect to $\lambda_2$ can be performed explicitly, since $Q(\lambda_1, \lambda_2)$ in (\ref{eq:G12A}), as a function of $\lambda_2$, is a pure Gaussian. This yields 
\bea \label{PDFr_2}
\tilde P(a_1,a_2) = \int \frac{d\lambda_1}{2\pi i} \, \,\frac {e^{\lambda_1 a_1}}{\sqrt{\cosh (\sqrt{2 \lambda_1})}} \frac{e^{-a_2^2/(4 C(\lambda_1))}}{\sqrt{4 \pi C(\lambda_1)}} \;,
\eea
where the integral is a Bromwich integral in the complex $\lambda_1$ plane. We thus insert this expression (\ref{PDFr_2}) in Eq. (\ref{PDFr_1}) and perform the integral over $a_1$ (the $\delta$ function in Eq. (\ref{PDFr_1}) enforcing $a_1 = z + a_2^2$ ) and we obtain

% \begin{widetext}
% \bea\label{PDFr_3}
% f_r(z) =  \int_{-\infty}^\infty da_2 \int \frac{d\lambda_1}{2\pi i} \, e^{\lambda_1 (z+a_2^2) - {a_2^2}/{(4 C(\lambda_1))}} \frac{1}{\sqrt{4\pi\,\cosh{(\sqrt{2\lambda_1})} \,C(\lambda_1)}  } \;.
% \eea
% \end{widetext}

\bea\label{PDFr_3}
f_r(z) =  \int_{-\infty}^\infty da_2 \int \frac{d\lambda_1}{2\pi i} \,  \frac{e^{\lambda_1 (z+a_2^2) - {a_2^2}/{(4 C(\lambda_1))}}}{\sqrt{4\pi\,\cosh{(\sqrt{2\lambda_1})} \,C(\lambda_1)}}. \;
\eea

The integral over $a_2$ can now be performed since this is a pure Gaussian integral (provided $\lambda_1 - 1/(4 C(\lambda_1)) < 0$, which can be achieved by moving the Bromwich contour) and we obtain
\bea\label{PDFr_4}
f_r(z) =   \int \frac{d\lambda_1}{2\pi i} \, e^{\lambda_1 z} \frac{1}{\sqrt{\cosh(\sqrt{2 \lambda_1})}} \frac{1}{\sqrt{1 - 4 \lambda_1 C(\lambda_1)}}.\;\;
\eea
Finally, using the expression for $C(\lambda_1)$ in Eq. (\ref{eq:C}), we have $1 - 4 \lambda_1 C(\lambda_1) = \tanh{(\sqrt{2\lambda_1})}/\sqrt{2 \lambda_1}$ and finally Eq. (\ref{PDFr_4}) can be written as (with the change of notation $\lambda_1 \to \lambda$)
\bea\label{PDFr_4}
f_r(z) =  \int \frac{d\lambda}{2\pi i} \, e^{\lambda z} \left[\frac{\sqrt{2 \lambda}}{\sinh{\sqrt{2 \lambda}}}\right]^{1/2},
\eea
which can equivalently be written as
\bea\label{PDFr_5}
\la e^{-\lambda z} \ra = \int_0^\infty e^{-\lambda\, z} f_r(z) \, dz = \left[\frac{\sqrt{2 \lambda}}{\sinh (\sqrt{2 \lambda})}\right]^{1/2} \;,
\eea
as stated in Eq. (13) in the main text. 

We now show how to perform the inverse Laplace transform in the r.h.s. of Eq. (\ref{PDFr_5}) to obtain an explicit expression for $f_r(z)$. We start by re-writing the r.h.s. as
\bea\label{LTinv1}
 \left[\frac{\sqrt{2 \lambda}}{\sinh (\sqrt{2 \lambda})}\right]^{1/2} = 2^{3/4} \lambda^{1/4} \frac{e^{-\sqrt{\frac{\lambda}{2}}}}{(1 - e^{-2 \sqrt{2\lambda}})^{1/2}} \;.
\eea 
Performing a formal series expansion of the r.h.s. of Eq. (\ref{LTinv1}), we obtain
\bea\label{LTinv2}
\left[\frac{\sqrt{2 \lambda}}{\sinh (\sqrt{2 \lambda})}\right]^{1/2} = 2^{3/4} \sum_{k=0}^\infty \frac{(2k)!}{(k!)^2\, 2^{2k}} \lambda^{1/4} e^{-\sqrt{\frac \lambda 2}(1+4k)}.\;\;\;
\eea
The idea is now to invert the Laplace transform term by term in the r.h.s. of (\ref{LTinv2}). Each term has the form 
$h(\lambda) = \lambda^{1/4}\, e^{-b \sqrt{\lambda}}$ with an appropriate $b$ that depends on $k$. Hence we need to find the inverse Laplace transform of $h(\lambda)$. This is not completely straightforward. To proceed, we first re-write $h(\lambda)$ as a product of two terms
\bea\label{eq:h1}
h(\lambda) = \frac{1}{\lambda^{1/4}} \sqrt{\lambda}\,e^{-b \sqrt{\lambda}} \;.
\eea	 
The first term $1/\lambda^{1/4}$ can be inverted easily using 
\bea \label{LTinv3}
{\cal L}^{-1}_{\lambda \to x} \left( \frac{1}{\lambda^{1/4}}  \right) = \frac{1}{\Gamma(1/4)\, x^{3/4}} \;,
\eea
and also the term $\sqrt{\lambda}\,e^{-b \sqrt{\lambda}}$ can be inverted using
\bea \label{LTinv4}
{\cal L}^{-1}_{\lambda \to x} \left( \sqrt{\lambda}\,e^{-b \sqrt{\lambda}} \right) = \frac{e^{-\frac{b^2}{4 x}} \left(b^2-2 x\right)}{4 \sqrt{\pi } x^{5/2}} \;.
\eea
The Laplace transform of the product can then be obtained by the convolution theorem and performing this convolution explicitly, we get
\begin{widetext}
 \bea \label{LTinv5}
{\cal L}^{-1}_{\lambda \to x} \left(  \lambda^{1/4}\, e^{-b \sqrt{\lambda}} \right) =  \frac{\sqrt{b}}{8\sqrt{2} \pi x^{5/2}} e^{-\frac{b^2}{8x}} \left[ (b^2-2x) K_{1/4}\left(\frac{b^2}{8x}\right) + b^2 K_{3/4}\left(\frac{b^2}{8x}\right) \right] \;,
\eea
\end{widetext}
where $K_{\nu}(z)$ is the modified Bessel function of index $\nu$. Using this result (\ref{LTinv5}) for each term in Eq. (\ref{LTinv2}) we get our final expression for the scaling function 
\begin{widetext}
 \begin{equation} \label{LTinv6}
f_r(z) = \frac{1}{4\pi z^{3/2}} \sum_{k=0}^\infty \frac{(2k)!}{(k!)^2\, 2^{2k}} \sqrt{4k+1} \, e^{-\frac{(4k+1)^2}{16z}} \left(\left(\frac{(4k+1)^2}{4z}-1 \right)K_{1/4}\left(\frac{(4k+1)^2}{16z} \right) + \frac{(4k+1)^2}{4z} K_{3/4}\left(\frac{(4k+1)^2}{16z} \right) \right) 
 \end{equation}
 \end{widetext}
Even though this expression is a bit long, it is straightforward to plot this function using Mathematica (indeed the series converges very fast), as shown in Fig. 2(b) in the main text.  

\vspace*{0.3cm}
\noindent{\it Asymptotic behaviours of $f_r(z)$:}  
%\subsection{Asymptotic behaviours of $f_r(z)$}
The limit $z \to 0$ is easy to obtain as it is given by the $k=0$ term of the series in Eq. (\ref{LTinv6}). Using $K_{\nu}(x) \simeq \sqrt{\pi/(2x)} e^{-x}$ as $x \to \infty$, we find that the leading order behavior of the $k=0$ term, and hence that of $f_r(z)$, is given by
\bea\label{fr_small_z}
f_r(z) \simeq \frac{1}{2\sqrt{2\pi}\,z^2} \, e^{-\frac{1}{8z}} \;, \quad {\textrm as} \quad z \to 0 \;. 
\eea
In contrast, the other limit $z\to \infty$ is trickier as in the case of $f_x(a_1)$ in Eq. (\ref{LTa1_4}). To proceed, it is convenient to go back to the original Laplace transform in Eq. (\ref{PDFr_5}) and re-write its r.h.s. using the following identity
\bea\label{id_sinh}
\frac{\sqrt{2\lambda}}{\sinh(\sqrt{2 \lambda})} = 2 \sum_{n=1}^\infty (-1)^{n+1} \frac{n^2 \pi^2}{n^2 \pi^2 + 2\lambda} \;.
\eea
Hence we get from Eq. (\ref{PDFr_5}) the following alternative representation of the Laplace transform which is more suited for the large $z$ analysis of $f_r(z)$
\bea\label{LTinv7}
\int_0^\infty e^{-\lambda z} f_r(z) \,dz = \left[ 2 \sum_{n=1}^\infty (-1)^{n+1} \frac{n^2 \pi^2}{n^2 \pi^2 + 2\lambda}\right]^{1/2}
\eea
For large $z$, the dominant contribution comes from the branch-cut around the pole corresponding to the $n=1$ term in this series in Eq.~(\ref{LTinv7}). Near this pole corresponding to $n=1$, the r.h.s. behaves as $\simeq \pi\sqrt{2}/\sqrt{\pi^2+2\lambda}$. Using the identity in Eq. (\ref{id_cosh3}) we obtain
\bea\label{fr_large}
f_r(z) \simeq \sqrt{\frac{\pi}{z}} \, e^{-\frac{\pi^2}{2}\,z} \quad {\textrm as} \quad z \to \infty \;.
\eea
The asymptotic behaviors of $f_r(z)$ in Eqs. (\ref{fr_small_z}) and (\ref{fr_large}) have been stated below Eq. (13) in the main text.

\end{document}